\documentclass[aps, showpacs, showkeys, nofootinbib, floatfix]{revtex4}
\usepackage{amssymb}
\usepackage{amsmath}
\usepackage{graphicx}

\begin{document}

\title{Constraints on a decomposed dark fluid with constant adiabatic sound speed by jointing the geometry test and growth rate after Planck data}

\author{Weiqiang Yang$^a$}
\author{Lixin Xu$^{a,}$$^b$\footnote{lxxu@dlut.edu.cn}}
\author{Yuting Wang$^c$}
\author{Yabo Wu$^d$}

\affiliation{$^a$Institute of Theoretical Physics, School of Physics and Optoelectronic Technology, Dalian University of Technology, Dalian 116024, People's Republic of China \\
$^b$College of Advanced Science and Technology, Dalian University of
Technology, Dalian 116024, People's Republic of China
\\
$^c$National Astronomy Observatories, Chinese Academy of Science, Beijing, 100012, People's Republic of China \\
$^d$Department of Physics, Liaoning Normal University, Dalian 116029, People's Republic of China
}

\begin{abstract}
In this paper, a unified dark fluid with constant adiabatic sound speed is decomposed into cold dark matter interacting with vacuum energy. Based on Markov chain Monte Carlo method, we constrain this model by jointing the geometry and dynamical measurement. The geometry test includes cosmic microwave background radiation from \textit{Planck}, baryon acoustic oscillation, and type Ia supernovae; the dynamic measurement is $f\sigma_8(z)$ data points which is obtained from the growth rate via redshift-space distortion, and $\sigma_8(z)$ is the root-mean-square amplitude of the density contrast $\delta$ at the comoving $8h^{-1}$ Mpc scale. The jointed constraint shows that $\alpha$ = $    0.000662_{-    0.000662}^{+    0.000173}$ and $\sigma_8$ = $    0.824_{-    0.0166}^{+    0.0128}$. The CMB and matter power spectra are both similar for the case of $\alpha=$ mean value and that of $\alpha=0$. However, the evolutionary curves of $f\sigma_8(z)$ are different. This means that, to some extent, the data points of the growth rate could break the degeneracy of the dark energy models.
\end{abstract}

\pacs{98.80.-k, 98.80.Es}
\maketitle

\section{Introduction}

Accelerating expansion of the Universe has been shown from the type Ia supernova (SNIa) observations \cite{ref:SN-Riess1998,ref:SN-Perlmuter1999}, cosmic microwave background (CMB) anisotropy measurement from Wilkinson Microwave Anisotropy Probe (WMAP) \cite{ref:CMB-Spergel2003}, and large scale structure from the Sloan Digital Sky Survey \cite{ref:LSS-Pope2004}. In order to explain this mechanism, theorists introduce an exotic energy component with negative pressure, which is called as dark energy. Based on this opinion, a lot of dark energy models have been proposed. One kind of popular  model is the unified dark fluid, these models have been put forward and further studied in Refs. \cite{ref:UDF1,ref:UDF2,ref:UDF3-Xu2012-diffcs2,ref:UDF4,ref:UDF5,ref:UDF6,ref:UDF7,ref:UDF8,ref:UDF9,ref:UDF10,ref:UDF11,ref:UDF12,ref:UDF13}.
In principle, the equation of state (EoS) can be determined up to an integration constant by the adiabatic sound speed. The model of zero adiabatic sound speed $c^2_s$ has been studied in Ref. \cite{ref:UDF11}; the case of time variable $c^2_s$ was discussed in Ref. \cite{ref:UDF12}; the model of constant adiabatic sound speed (CASS)-that is,$c^2_s=\alpha$-has been studied in Refs. \cite{ref:CASS1-Balbi2007,ref:CASS2-Xu2012}. In Ref. \cite{ref:CASS2-Xu2012}, Xu et al. took the CASS model as a whole dark fluid, and found that small values of $\alpha$ are favored by using the Markov Chain Monte Carlo (MCMC) method with SNIa Union 2 \cite{ref:SN557-Suzuki2012}, baryon acoustic oscillation (BAO) \cite{ref:BAO-Percival2010}, and the full CMB information from seven-year WMAP data sets \cite{ref:WMAP7-Komatsu2011}.

Recently, it has been shown that any unified dark fluid model can be decomposed into pressureless dark matter interacting with a vacuum energy \cite{ref:vacuum1-Wands2012,ref:vacuum2-De-Santiago2012,ref:vacuum3-GCG-Wang2013}. Following these papers, we want to study the decomposed CASS model described by cold dark matter interacting with a vacuum energy. Thus, we call the model as IDCASS (interacting decomposed dark fluid model with constant adiabatic sound speed). The interacting dark energy model can introduce some new features to the structure formation, as one can see in Refs. \cite{ref:IDEgrowth1-CC2009,ref:IDEgrowth2-Song2009,ref:IDEgrowth3-Baldi2011,
ref:IDEgrowth4-Hellwing2010,ref:IDEgrowth5-Rubano2011}. In order to explore the possibility of interacting dark energy, it is necessary to consider the effect of the interaction on the structure formation.

As for the observational aspect, to break the possible degeneracy of cosmological models, the geometry information (SNIa, BAO, CMB) is not enough because the different models may undergo similar background evolution behavior, but the dynamical growth history could be different. Therefore, the large scale structure information is a powerful tool to discriminate the dark energy models. Via the redshift-space distortion (RSD), the measurement of the growth rate $f$ is closely related to the evolutionary speed of matter density contrast $\delta$, where $f=d\ln\delta/d\ln a$. It is worth to notice that the growth rate $f$ has been used to constrain the dark energy model and to test the growth index in Refs. \cite{ref:f1-Nesseris2008,ref:f2-Basilakos2012,ref:f3-Gupta2012,ref:f4-Paul2012,ref:f5-Shi2012}. However, considering the cosmological constant and cold dark matter ($\Lambda$CDM) model, the observational values of the growth rate $f_{obs}=\beta b$ are derived from the redshift-space distortion parameter $\beta$ and the linear bias $b$. It means that the current $f_{obs}$ data is model-dependent and not suitable to constrain the other models. To avoid this issue, $f\sigma_8(z)$ will provide a good test of dark energy models ($\sigma_8(z)$ is the root-mean-square mass fluctuation in spheres with radius $8h^{-1}$ Mpc). The model-independent measurement $f\sigma_8(z)$ is firstly proposed to constrain the dark energy models in Ref. \cite{ref:fsigma8-DE-Song2009}. Then, in Ref. \cite{ref:fsigma8-HDE-Xu2013}, Xu combined the geometry test with $f\sigma_8(z)$ data to constrain the holographic dark energy model and obtained a tight constraint of holographic parameter $c$. Furthermore, after \textit{Planck}, Xu parameterized the growth function as $f=\Omega_m^{\gamma_L}$ and compared the deviation of the growth index $\gamma_L$ in the Einstein's gravity theory and modified gravity theory in Ref. \cite{ref:fsigma8andPlanck-MG-Xu2013}.

The observational data points of $f\sigma_8(z)$ were provided by the 2dFGRS \cite{ref:fsigma81-Percival2004}, WiggleZ \cite{ref:fsigma82-Blake2011}, SDSS LRG \cite{ref:fsigma83-Samushia2012}, BOSS \cite{ref:fsigma84-Reid2012}, 6dFGRS \cite{ref:fsigma85-Beutler2012}, and VIPERS \cite{ref:fsigma86-Torre2013}. The former nine data points were summarized in Table 1 of Ref. \cite{ref:fsigma8total-Samushia2013}. The data point at $z=0.8$ was recently released by VIPERS in Ref. \cite{ref:fsigma86-Torre2013}. A lower growth rate from RSD than expected from \textit{Planck} was also pointed out in Ref. \cite{ref:fsigma87-Macaulay2013}. In this paper, the ten data points are shown in Table \ref{tab:fsigma8data}.

\begin{table}
\begin{center}
\begin{tabular}{ccc}
\hline\hline z & $f\sigma_8(z)$ & Survey and Refs \\ \hline
$0.067$ & $0.42\pm0.06$ & $6dFGRS~(2012)$ \cite{ref:fsigma85-Beutler2012}\\
$0.17$ & $0.51\pm0.06$ & $2dFGRS~(2004)$ \cite{ref:fsigma81-Percival2004}\\
$0.22$ & $0.42\pm0.07$ & $WiggleZ~(2011)$ \cite{ref:fsigma82-Blake2011}\\
$0.25$ & $0.39\pm0.05$ & $SDSS~LRG~(2011)$ \cite{ref:fsigma83-Samushia2012}\\
$0.37$ & $0.43\pm0.04$ & $SDSS~LRG~(2011)$ \cite{ref:fsigma83-Samushia2012}\\
$0.41$ & $0.45\pm0.04$ & $WiggleZ~(2011)$ \cite{ref:fsigma82-Blake2011}\\
$0.57$ & $0.43\pm0.03$ & $BOSS~CMASS~(2012)$ \cite{ref:fsigma84-Reid2012}\\
$0.60$ & $0.43\pm0.04$ & $WiggleZ~(2011)$ \cite{ref:fsigma82-Blake2011}\\
$0.78$ & $0.38\pm0.04$ & $WiggleZ~(2011)$ \cite{ref:fsigma82-Blake2011}\\
$0.80$ & $0.47\pm0.08$ & $VIPERS~(2013)$ \cite{ref:fsigma86-Torre2013}\\
\hline\hline
\end{tabular}
\caption{The data points of $f\sigma_8(z)$ measured from RSD with the survey references.}
\label{tab:fsigma8data}
\end{center}
\end{table}

The paper is organized as follows. In Sec. II, we revisit the unified dark fluid with constant adiabatic sound speed and decompose it into pressureless dark matter interacting with a vacuum energy. In Sec. III, we give the first-order and second-order perturbation equations of the cold dark matter and baryon, respectively, and obtain the evolution equations of growth rate for these two components. In Sec. IV, by adopting the MCMC method with the cosmic observational data sets, we show the model parameter space. Section V is the summary.

\section{A DECOMPOSED UNIFIED DARK FLUID WITH CONSTANT ADIABATIC SOUND SPEED}

Following Refs. \cite{ref:CASS1-Balbi2007} and \cite{ref:CASS2-Xu2012}, we consider a unified dark fluid with constant adiabatic sound speed
\begin{eqnarray}
c^2_s=\alpha.
\label{eq:CASS}
\end{eqnarray}

The definition of the CASS model tells us that $\alpha>0$ because the adiabatic sound speed is positive. After an integration, the EoS of the unified dark fluid can be determined
\begin{eqnarray}
w_u=\alpha-\frac{A}{\rho_u},
\label{eq:EoS}
\end{eqnarray}
where A is an integration constant. When $\alpha=0$ happens, Eq. (\ref{eq:EoS}) shows $w_u=-A/\rho_u$, which looks like the Chaplygin gas model, meanwhile, the sound speed is zero, so this is not a reasonable model.  If $A=0$ is assumed, Eq. (\ref{eq:EoS}) becomes $w_u=\alpha$, and the total EoS of dark fluid is a constant, which looks like a quintessence model. However, if the CASS model is taken as a unified dark sector, it would not happen, because it would look like a combination of cold dark matter and a simple cosmological constant.

The energy conservation equation for the dark fluid is
\begin{eqnarray}
\dot{\rho_u}+3H(\rho_u+P_u)=0.
\label{eq:udf-con-eq}
\end{eqnarray}

Combining Eqs. (\ref{eq:EoS}) and (\ref{eq:udf-con-eq}), the energy density of unified dark fluid can be written as
\begin{eqnarray}
{\rho _{u}} = {\rho _{u0}}\left[ {\left( {1 - {B_s}} \right) + {B_s}{a^{ - 3(1 + \alpha )}}} \right],
\label{eq:rho-udf}
\end{eqnarray}
where $\rho_{u0}$ is the present value of energy density, and $B_s=1-A/(\rho_{u0}(1+\alpha))$.

In a spatially flat universe, the Friedmann-Robertson-Walker metric reads
\begin{eqnarray}
{\rm{d}}{s^2} =  - d{t^2} + {a^2}(t)[d{r^2} + {r^2}(d{\theta ^2} + {\sin ^2}d{\varphi ^2})],
\label{eq:FRW}
\end{eqnarray}
and one can obtain the Friedmann equation \cite{ref:CASS2-Xu2012}
\begin{eqnarray}
{H^2} = H_0^2\left\{ {{\Omega _b}{a^{ - 3}} + {\Omega _r}{a^{ - 4}} + \left( {1 - {\Omega _b} - {\Omega _r}} \right)\left[ {(1 - {B_s}) + {B_s}{a^{ - 3(1 + \alpha )}}} \right]} \right\},
\label{eq:Friedmann}
\end{eqnarray}
where $\Omega_i (i=b,r)$ are dimensionless energy parameters of the baryon and radiation, respectively.

In Refs. \cite{ref:vacuum1-Wands2012,ref:vacuum2-De-Santiago2012,ref:vacuum3-GCG-Wang2013}, a unified dark fluid can be decomposed into pressureless dark matter interacting with a vacuum energy, of course. For the CASS model, we decompose it as
\begin{eqnarray}
\rho_u=\rho_c+V,
\label{eq:decompose-rho-udf}
\end{eqnarray}
where $\rho_c$ and $V$ are, respectively, the energy density of cold dark matter and vacuum energy. For the decomposed model, the Friedmann equation can be written as
\begin{eqnarray}
{H^2} = H_0^2\left\{ {{\Omega _b}{a^{ - 3}} + {\Omega _r}{a^{ - 4}} + {\Omega _V}{\rm{ + }}{\Omega _c}\frac{{\alpha  + {a^{ - 3(1 + \alpha )}}}}{{1 + \alpha }}} \right\},
\label{eq:decompose-Friedmann}
\end{eqnarray}
where $\Omega_i (i=c,V)$ are dimensionless energy parameters of cold dark matter and vacuum energy, respectively. So the IDCASS model only has one degree of freedom $\alpha$ whereas the original CASS model is characterised by two model parameters, $\alpha$ and A (or $B_s$).

The energy conservation equations of cold dark matter and vacuum energy are
\begin{eqnarray}
\dot{\rho_c}+3H\rho_c=-Q,
\label{eq:int-cdm-con-eq}
\end{eqnarray}
\begin{eqnarray}
\dot{V}=Q,
\label{eq:int-vacuum-con-eq}
\end{eqnarray}
where $Q$ is the energy transfer between dark matter and vacuum energy. Combining the above two equations with Eqs. (\ref{eq:EoS}) and (\ref{eq:decompose-rho-udf}), we obtain
\begin{eqnarray}
Q=3\alpha H\rho_c.
\label{eq:interaction}
\end{eqnarray}

Here, it is necessary to say that the way of decomposing the unified dark fluid is not unique, but Eq. (\ref{eq:decompose-rho-udf}) is a convenient choice and does not introduce some other degrees of freedom. In this decomposed case, the model has only one degree of freedom $\alpha$. Before the decomposition, the model is taken as a whole dark fluid, the model parameter $\alpha$ not only represents the adiabatic sound speed, but also influences the EoS together with the other parameter $A$. After the decomposition, apart from describing the sound speed and affecting the total EoS of dark matter and dark energy, $\alpha$ will reveal some possible characters inside the dark sectors. Concretely, this parameter would show the interacting intensity between dark matter and vacuum energy, and change the evolution of the effective EoS for the two dark components. Besides, the interaction also affects the evolution of perturbation equations of dark matter.

\section{The PERTURBATION EQUATIONS AND GROWTH RATE}

We consider the scalar perturbations in a spatially flat Universe, whose line element is \cite{ref:ds1-Mukhanov1992,ref:ds2-Malik,ref:ds3-Malik2009,ref:ds4-Malik2005}
\begin{eqnarray}
d{s^2} =  - (1 + 2\phi )d{t^2} + 2a{\partial _i}Bdtd{x^i} + {a^2}\left[ {(1 - 2\psi ){\delta _{ij}} + 2{\partial _i}{\partial _j}E} \right]d{x^i}d{x^j}.
\label{eq:ds2-perturbed}
\end{eqnarray}

In the general case of interacting fluids, the covariant conservation equation of fluid A reads
\begin{eqnarray}
\nabla_{\mu}T^{\mu \nu}_A=Q^{\mu}_A.
\label{eq:covariant-con-eq}
\end{eqnarray}

In Refs. \cite{ref:ds2-Malik,ref:ds3-Malik2009,ref:ds4-Malik2005,ref:Q3-Kodama1984,ref:Q4-Malik2003}, the perturbed energy-momentum transfer can be split into
\begin{eqnarray}
Q^A_{\mu}=[-Q_A(1+\phi)-\delta Q_A,\partial_i(F_A+Q_A\theta)].
\label{eq:convariant-interaction}
\end{eqnarray}

The energy and momentum conservation equations for fluid A become \cite{ref:ds2-Malik,ref:ds3-Malik2009,ref:ds4-Malik2005,ref:Q3-Kodama1984,ref:Q4-Malik2003}
\begin{eqnarray}
\dot{\delta {\rho}}_A + 3H\left( {\delta {\rho _A} + \delta {P_A}} \right) - 3\left( {{\rho _A} + {P_A}} \right)\dot \psi  + \left( {{\rho _A} + {P_A}} \right)\frac{{{\nabla ^2}}}{{{a^2}}}\left( {{\theta _{\rm{A}}} + \sigma } \right) = \delta {Q_A} + {Q_A}\phi,
\label{eq:general-density-per}
\end{eqnarray}
\begin{eqnarray}
\left( {{\rho _A} + {P_A}} \right){{\dot \theta }_A} - 3c_{sA}^2H\left( {{\rho _A} + {P_A}} \right){\theta _{\rm{A}}} + \left( {{\rho _A} + {P_A}} \right)\phi  + \delta {P_A} + \frac{2}{3}\frac{{{\nabla ^2}}}{{{a^2}}}{\Pi _A} = {F_A}{\rm{ + }}{Q_A}\theta  - \left( {1 + c_{sA}^2} \right){Q_A}{\theta _A}.
\label{eq:general-velocity-per}
\end{eqnarray}

In the synchronous gauge, the perturbation equations for baryon density contrast and velocity are
\begin{eqnarray}
\dot{\delta}_b-\frac{k^2}{a^2}\theta_b=-\frac{\dot{h}}{2},
\label{eq:baryon-1nd-density-per}
\end{eqnarray}
\begin{eqnarray}
\dot{\theta}_b=0.
\label{eq:baryon-velociry}
\end{eqnarray}

Using the relation expression $\ddot{h}+2H\dot{h}=-8\pi G(\delta\rho+3\delta P)$, we can obtain the second-order differential equations for the baryon density contrast
\begin{eqnarray}
\ddot{\delta}_b+2H\dot{\delta}_b=4\pi G(\delta\rho+3\delta P).
\label{eq:baryon-2nd-density-per}
\end{eqnarray}

In the synchronous gauge, considering the geodesic case of the interacting vacuum energy and dark matter model in Refs. \cite{ref:vacuum2-De-Santiago2012,ref:vacuum3-GCG-Wang2013}, we introduce an energy flow that is parallel to the four-velocity of the dark matter $Q^{\mu}_c=-Q u^{\mu}_c$. In this case, $F_c-Q(\theta-\theta_c)=0$ in Eq. (\ref{eq:convariant-interaction}) \cite{ref:parallelQ1-Valiviita2008,ref:parallelQ2-Koyama2009}, and the velocity perturbation for dark matter is zero. So the first-order and second-order differential equations for the dark matter density contrast can be derived \cite{ref:vacuum3-GCG-Wang2013}
\begin{eqnarray}
\dot{\delta}_c=-\frac{\dot{h}}{2}+\frac{Q}{\rho_c}\delta_c,
\label{eq:cdm-1nd-density-per}
\end{eqnarray}
\begin{eqnarray}
{{\ddot \delta }_c} + \left( { - \frac{Q}{{{\rho _c}}} + 2H} \right){{\dot \delta }_c} - \left[ {2H\frac{Q}{{{\rho _c}}} + \dot{\left( {\frac{Q}{{{\rho _c}}}} \right)}} \right]{\delta _c} = 4\pi G\left( {\delta \rho  + \delta P} \right).
\label{eq:cdm-2nd-density-per}
\end{eqnarray}

According to Refs. \cite{ref:f1-Nesseris2008,ref:g1-Wang1998}, the growth factor $g(a)$ is proportional to the linear density perturbation $\delta=\delta\rho/\rho$, $g(a)=\delta(a)/a$, one can obtain the growth factor for the dark matter and baryon
\begin{eqnarray}
\frac{{{d^2}{g_c}}}{{d\ln {a^2}}} + \left[ {\frac{5}{2} - 3\alpha  + \frac{3}{2}w_{eff}(a){\Omega _V}(a)} \right]\frac{{d{g_c}}}{{d\ln a}}{\rm{ + }}\frac{3}{2}\left( {{\rm{1}} - 3\alpha } \right)\left[ {1 + w_{eff}(a){\Omega _V}(a)} \right]{g_c} =  \frac{3}{2}\left[ {{\Omega _c}(a){g_c} + {\Omega _b}(a){g_b}} \right] ,
\label{eq:cdm-g}
\end{eqnarray}
\begin{eqnarray}
\frac{{{d^2}{g_b}}}{{d\ln {a^2}}} + \left[ {\frac{5}{2} + \frac{3}{2}w_{eff}(a){\Omega _V}(a)} \right]\frac{{d{g_b}}}{{d\ln a}}{\rm{ + }}\frac{3}{2}\left[ {1 + w_{eff}(a){\Omega _V}(a)} \right]{g_b} = \frac{3}{2}\left[ {{\Omega _c}(a){g_c} + {\Omega _b}(a){g_b}} \right],
\label{eq:bar-g}
\end{eqnarray}
where
\begin{eqnarray}
\Omega_c(a)=\frac{H^2_0}{H^2}\Omega_{c0}a^{-3(1+\alpha)},~~~
\Omega_b(a)=\frac{H^2_0}{H^2}\Omega_{b0}a^{-3}.
\label{eq:Omega_c,b}
\end{eqnarray}

Here, $w_{eff}(a)$ is the effective EoS of dark energy which is defined as $w_{eff}(a) \equiv \frac{1}{lna} \int^{lna}_0 d\ln a' w_V(a')$ in Ref. \cite{ref:WMAP5}, $D(a)$ is the growth ratio of perturbation amplitude at some scale factor relative to the normalized scale factor, whose relationship with $f(a)$ is $f=d\ln D/d\ln a=d\ln\delta/d\ln a$. The right-hand side of Eqs. (\ref{eq:cdm-g}) and (\ref{eq:bar-g}) is the cross term between these two equations, so if we want to know $g_c$ or $g_b$, we need to solve the equation set. Moreover, according to $g_m=\rho_cg_c/(\rho_c+\rho_b)+\rho_bg_b/(\rho_c+\rho_b)$, we can obtain the growth factor of the matter.

\section{COSMOLOGICAL IMPLICATIONS AND CONSTRAINTS}

\subsection{Implications on CMB temperature and matter power spectra for the model parameter $\alpha$}

Here, we illustrate how the CMB temperature and matter power spectra are characterized by different values of the model parameter $\alpha$.

\begin{figure}[!htbp]
\includegraphics[width=13cm,height=9cm]{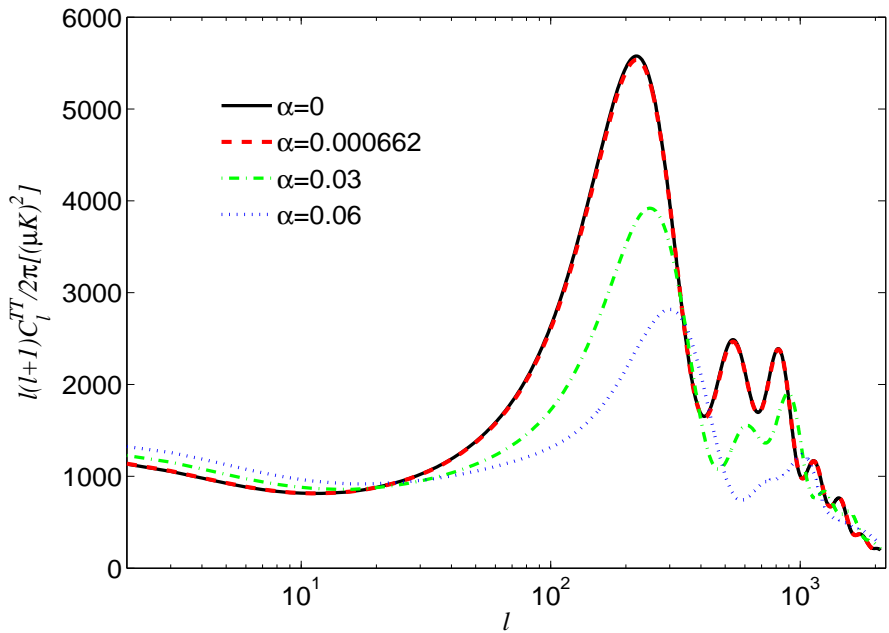}
  \caption{The effects on CMB temperature power spectra for the different values of model parameter $\alpha$. The black solid, red thick dashed, green dotted-dashed, and blue dotted lines are for $\alpha=0, 0.000662, 0.03$, and $0.06$, respectively; the other relevant parameters are fixed with the mean values as shown in the fourth column of Table \ref{tab:total-mean-table}.}
  \label{fig:CMBpower_tot}
\end{figure}

\begin{figure}[!htbp]
\includegraphics[width=11cm,height=8cm]{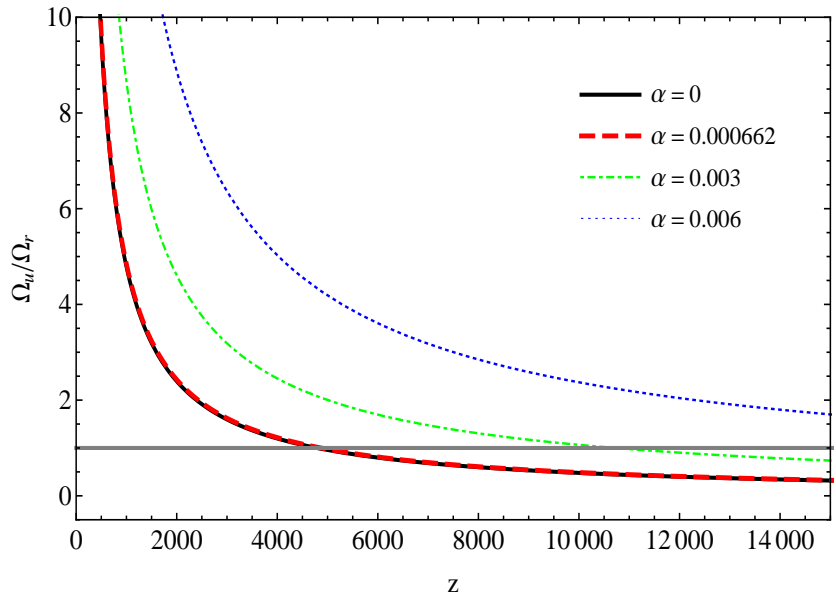}
  \caption{The evolutionary curves for the ratio of dark fluid and radiation $\Omega_u/\Omega_r$ when the parameter $\alpha$ is varied. The different lines correspond to the cases in Fig. \ref{fig:CMBpower_tot}; the horizontal gray thick line corresponds to the case of $\Omega_u=\Omega_r$, and the other relevant parameters are fixed with the mean values as shown in the fourth column of Table \ref{tab:total-mean-table}.}
  \label{fig:Omega_u-r_tot}
\end{figure}

First, the effects on the CMB temperature power spectra are shown in Fig. \ref{fig:CMBpower_tot}. At the same time, in order to clearly explain the change of CMB power spectra, we also plot the evolutionary curves for the ratio of dark fluid and radiation $\Omega_u/\Omega_r$ in Fig. \ref{fig:Omega_u-r_tot}, for which $\Omega_u=\Omega_c$ in the early epoch. Following the discussion of Ref. \cite{ref:Hu2001}, increasing the value of $\alpha$, which is equivalent to increasing the value of the effective dimensionless energy density of cold dark matter $\Omega_c$, will make the equality of matter and radiation occur earlier. This raises $l_{eq}$ (the horizon scale at matter-rediation equality) and reduces the the driving effect that the decay of the gravitational potential happens on the acoustic oscillations during the radiation era. As a result, the first peak of CMB power spectra is depressed. Moreover, since the parameter $\alpha$ has an effect on the expansion rate, the angular diameter distance to recombination becomes larger when $\alpha$ increases, which makes the positions of peaks shift towards the right side. As is shown in Eq. (\ref{eq:Omega_c,b}), the values of $\alpha$ describe the possible deviation from the standard evolution scaling law $a^{-3}$ of effective dark matter. At large scales $l<100$, the varied parameter $\alpha$ affects the CMB power spectra via Integrated Sachs-Wolfe (ISW) effect due to the evolution of gravitational potential. Via changing the expansion history of the Universe, the ISW effect on CMB power spectra has been studied in Ref. \cite{ref:Das01}. Moveover, in comparison with changing the primordial power spectra \cite{ref:Das02}, the ISW effect does not affect the polarization power spectra and hence CMB polarization spectra at low multiples could in principle be used to distinguish the effect from power deficit originating features in the primordial power spectra.

\begin{figure}[!htbp]
\includegraphics[width=13cm,height=9cm]{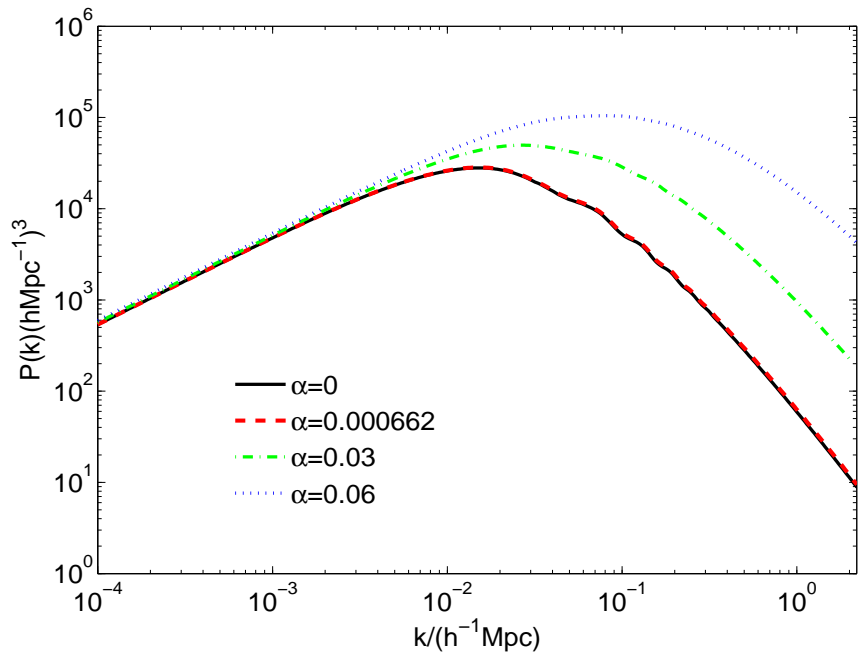}
  \caption{The effects on matter power spectra for the different values of model parameter $\alpha$. The black solid, red thick dashed, green dotted-dashed, and blue dotted lines are for $\alpha=0, 0.000662, 0.03$, and $0.06$, respectively; the other relevant parameters are fixed with the mean values as shown in the fourth column of Table \ref{tab:total-mean-table}.}
  \label{fig:Mpower_tot}
\end{figure}

Then, in Fig. \ref{fig:Mpower_tot}, we plot the matter power spectrum $P(k)$ when we use $f\sigma_8$ data. For $\alpha>0$ (the energy transfer is from dark matter to vacuum energy), with the increasing of $\alpha$, $P(k)$ is enhanced in the small-scale due to the earlier matter-radiation equality which moves the turnover in the matter power spectrum to smaller scales.

From the CMB and matter power spectra, it is easy to see that the case of $\alpha=$ mean value ($\alpha=0.000662$) and that of $\alpha=0$ (corresponding to the $\Lambda$CDM model) are very similar, so it is difficult to distinguish the IDCASS model from the $\Lambda$CDM model. However, due to using the $f\sigma_8(z)$ data set of large scale structure information, we hope that the different dynamical growth history could break the degeneracy of the models.

\subsection{The growth rate after \textit{Planck} for the interacting decomposed dark fluid with constant adiabatic sound speed}

\begin{figure}[!htbp]
\includegraphics[width=13cm,height=9cm]{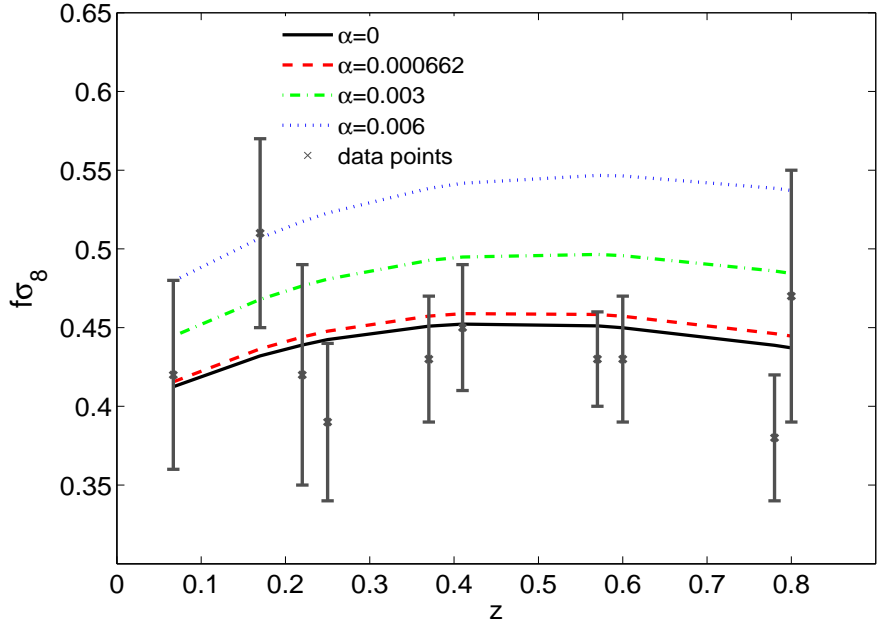}
  \caption{The fitting evolutionary curves of $f\sigma_8(z)$ about the redshift $z$ for the varied model parameter $\alpha$. The black solid, red dashed, green dotted-dashed, and blue dotted lines are for $\alpha=0, 0.000662$, $0.003$, and $0.006$, respectively; The gray error bars denote the observations of $f\sigma_8$ are listed in Table \ref{tab:fsigma8data}; the other relevant parameters are fixed with the mean values as shown in the fourth column of Table \ref{tab:total-mean-table}, when we use the $f\sigma_8(z)$.}
  \label{fig:fsigma8_tot}
\end{figure}

In order to test the effects on evolutions of $f\sigma_8(z)$ for the model parameter $\alpha$, we fix the relevant cosmological parameters according to the fourth column of Table \ref{tab:total-mean-table} but consider $\alpha$ to be varied in a range. The evolutionary curves of $f\sigma_8(z)$ with respect to the redshift z are shown in Fig. \ref{fig:fsigma8_tot}. With the increasing the values of $\alpha$, the curves of $f\sigma_8(z)$ are enhanced at both lower and higher redshifts.

Importantly, one can clearly see that the case of $\alpha=$ mean value ($\alpha=0.000662$) and that of $\alpha=0$ (correspond to the $\Lambda$CDM model) are distinguishing from the evolutionary curves of $f\sigma_8$, which is different from the evolutionary curves of CMB temperature and matter power spectra. It means that, to some extent, the growth rate (or $f\sigma_8$) data set could break the degeneracy between the IDCASS model and the $\Lambda$CDM model.

\subsection{Data sets and results}

In March 2013, the European Space Agency (ESA) and the \textit{Planck} Collaboration publicly released the new CMB data \cite{ref:Planck2013,ref:Planck2013-download} which are expected to improve the cosmological constraint. Here, we adopt the cosmic observational data sets which include SNIa SNLS3 \cite{ref:SNLS3-1,ref:SNLS3-2}, BAO (Sec. 5.2 of Ref. \cite{ref:Planck2013-params}), and \textit{Planck} data \cite{ref:Planck2013,ref:Planck2013-download}. The \textit{Planck} data sets which include two main parts: one is the high-l TT likelihood (\textit{CAMSpec}) up to a maximum multipole number of $l_{max}=2500$ from $l=50$; the other is the low-l TT likelihood up to $l=49$ and the low-l TE, EE, BB likelihood up to $l=32$ from WMAP nine-year data sets \cite{ref:WMAP9}. For more detailed descriptions about the cosmic observations, one can see Ref. \cite{ref:fsigma8andPlanck-MG-Xu2013}.

The seven-dimensional parameter space for the IDCASS model is
\begin{eqnarray}
P\equiv\{\Omega_bh^2, \Omega_ch^2, \Theta_S, \tau, \alpha, n_s, log[10^{10}A_S]\},
\label{eq:parameter_space}
\end{eqnarray}
where $\Omega_bh^2$ and $\Omega_ch^2$ stand for the density of the baryon and cold dark matter, respectively, $\Theta_S$ refers to the ratio of sound horizon and angular diameter distance, $\tau$ indicates the optical depth, $\alpha$ are the added parameters for the decomposed model, $n_s$ is the scalar spectral index, and $A_s$ represents the amplitude of the initial power spectrum. The pivot scale of the initial scalar power spectrum $k_{s0}=0.05Mpc^{-1}$ is used in this paper. The following priors to model parameters are adopted: $\Omega_bh^2\in[0.005,0.1]$, $\Omega_ch^2\in[0.01,0.99]$, $\Theta_S\in[0.5,10]$, $\tau\in[0.01,0.8]$, $\alpha\in[0,1]$, $n_s\in[0.5,1.5]$, $log[10^{10}A_S]\in[2.7,4]$. In order to obtain the model parameter space from the cosmic observations, we use the MCMC method and modify the publicly available \textbf{COSMOMC} \cite{ref:cosmomc-Lewis2002,ref:camb} package in which a new module was added to calculate the value of $f\sigma_8(z)$, one also can see Refs. \cite{ref:fsigma8-HDE-Xu2013,ref:fsigma8andPlanck-MG-Xu2013}.

In our numerical calculations, the total likelihood $\chi^2$ can be constructed as
\begin{eqnarray}
\chi^2=\chi^2_{CMB}+\chi^2_{BAO}+\chi^2_{SN}+\chi^2_{RSD}.
\label{eq:chi2}
\end{eqnarray}

\begin{figure}[!htbp]
\includegraphics[width=20cm,height=15.5cm]{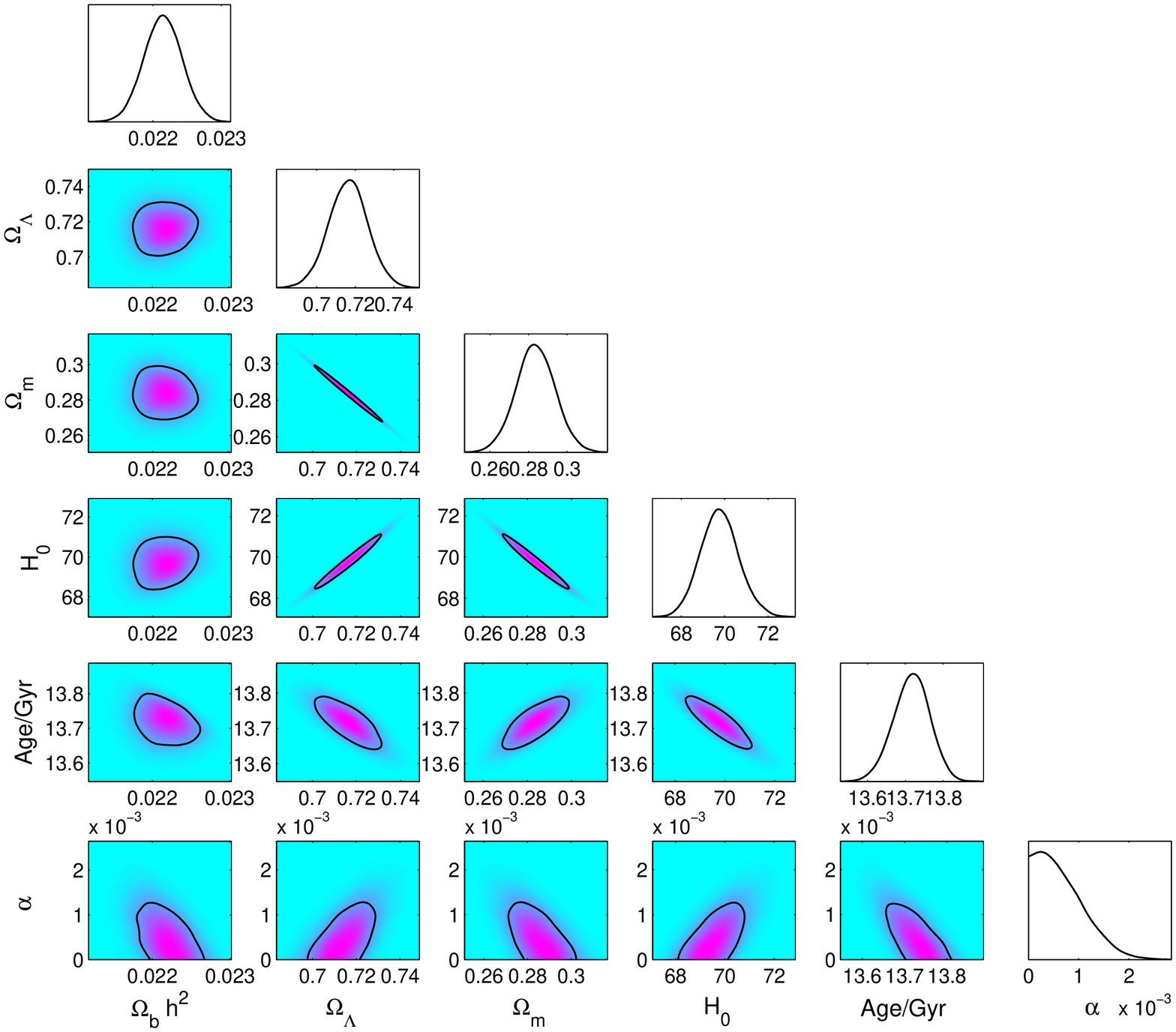}
  \caption{The 1D marginalized distributions on individual parameters and 2D contours with 68\%C.L.(confidence levels) between each other using the combination of the observational data points from the \textit{Planck} information, BAO, and SNIa, and RSD data sets.}
  \label{fig:contour_tot}
\end{figure}

We have run eight chains in parallel on the computer and checked the convergence to stop sampling when the worst e-values (the variance/mean or mean/variance) of 1/2 chains $R-1$ is of the order 0.01. When$f\sigma_8(z)$ is adopted, the constraint results are presented in the fourth column of Table \ref{tab:total-mean-table} and Fig. \ref{fig:contour_tot}. In the fourth column of Table \ref{tab:total-mean-table}, we list the mean values of basic and derived model parameters with $1\sigma, 2\sigma$, and $3\sigma$ regions. Then, in Fig. \ref{fig:contour_tot}, we show the one-dimensional (1D) marginalized distributions of parameters and two-dimensional (2D) contours with the confidence level. Moreover, in order to clearly see the effect on the cosmological constraint for the $f\sigma_8(z)$ data, we also constrain the decomposed model without $f\sigma_8(z)$ data set, the results are shown in the second column of Table \ref{tab:total-mean-table}.

The constraint results from \textit{Planck}, BAO, SNIa, and RSD data sets favor small intensity of interaction which is up to the order of $10^{-4}$, and the results without RSD data set show $\alpha=0.00159$. Obviously, the constraint with RSD data set is tighter than that without RSD data set, which means that the $f\sigma_8(z)$ data can improve the cosmological constraint results. The result for the parameter $\alpha$ is very similar to $\alpha=0.000487$ for the mean value in Table 1 of Ref. \cite{ref:CASS2-Xu2012}. However, this work is different from Ref. \cite{ref:CASS2-Xu2012} in the following several aspects. First, due to the recently released \textit{Planck} data, the high-precision data sets make the constraint results more reliable than WMAP seven-year data. Then, in this paper, the CASS model is not taken as a whole dark fluid, but considered as a decomposed fluid which include cold dark matter interacting with vacuum energy, and the IDCASS model has just one degree of freedom. Based on the decomposed model, it is natural to deduce an interaction form which is relevant to the model parameter $\alpha$. This expression allows us to explore the effects on the cosmic evolution from interacting dark energy. The last but most important aspect is adopting the large scale structure information ($f\sigma_8(z)$ from RSD), the dynamical evolution is powerful tool to break the possible degeneracy of some cosmological models. It means that the different dark energy models could have the same background evolution history, but the dynamical evolution would be different.

\begingroup
\squeezetable
\begin{center}
\begin{table}
\begin{tabular}{ccccc}
\hline\hline
Model Parameters & Mean value without $f\sigma_8(z)$ & Best fit without $f\sigma_8(z)$ & Mean value with $f\sigma_8(z)$ & Best fit with $f\sigma_8(z)$\\ \hline

$\Omega_b h^2$ & $    0.0218_{-    0.000310}^{+    0.000307}$ & $0.0220$ & $    0.0221_{-    0.000270}^{+    0.000268}$ & $0.0223$\\

$\Omega_c h^2$ & $    0.116_{-    0.00163}^{+    0.00165}$ & $0.117$ & $    0.115_{-    0.00157}^{+    0.00155}$ & $0.116$\\

$100\theta_{MC}$ & $    1.0414_{-    0.000580}^{+    0.000576}$ & $1.0414$ & $    1.0417_{-    0.000574}^{+    0.000580}$ & $1.0414$\\

$\tau$ & $    0.0884_{-    0.0128}^{+    0.0126}$ & $0.0830$ & $    0.0821_{-    0.0134}^{+    0.0115}$ & $0.0850$\\

$n_s$ & $    0.962_{-    0.00578}^{+    0.00578}$ & $0.963$ & $    0.967_{-    0.00566}^{+    0.00567}$ & $0.967$\\

${\rm{ln}}(10^{10} A_s)$ & $    3.0895_{-    0.0244}^{+    0.0241}$ & $3.0826$ & $    3.0670_{-    0.0252}^{+    0.0223}$ & $3.0744$\\ \hline

$\Omega_\Lambda$ & $    0.717_{-    0.0110}^{+    0.0110}$ & $0.712$ & $    0.716_{-    0.00983}^{+    0.00969}$ & $0.710$\\

$\Omega_m$ & $    0.283_{-    0.0110}^{+    0.0110}$ & $0.288$ & $    0.284_{-    0.00969}^{+    0.00983}$ & $0.290$\\

$\sigma_8$ & $  -  $ & $  -  $ & $    0.824_{-    0.0166}^{+    0.0128}$& $0.817$\\

$z_{re}$ & $   10.978_{-    1.0919}^{+    1.0788}$ & $10.497$ & $   10.285_{-    1.0479}^{+    1.0507}$ & $10.524$\\

$H_0$ & $   69.986_{-    0.984}^{+    0.974}$ & $69.597$ & $   69.769_{-    0.860}^{+    0.834}$ & $69.216$\\

${\rm{Age}}/{\rm{Gyr}}$ & $   13.673_{-    0.0575}^{+    0.0586}$ & $13.683$ & $   13.716_{-    0.0457}^{+    0.0503}$ & $13.746$\\

$\alpha$ & $    0.00159_{-    0.000802}^{+    0.000724}$ & $0.00139$ & $    0.000662_{-    0.000662}^{+    0.000173}$ & $0.000263$\\

\hline\hline
\end{tabular}
\caption{The constraint results of basic and derived model parameters with $1\sigma$ region from the cosmic observations. The mean and best fit values in the second and third columns are from the \textit{Planck} information, BAO, and SNIa data sets; the mean and best fit values in the fourth and fifth columns are from the \textit{Planck} information, BAO, SNIa, and $f\sigma_8(z)$ (RSD) data sets.}
\label{tab:total-mean-table}
\end{table}
\end{center}
\endgroup

\section{SUMMARY}

The unified dark fluid with constant adiabatic sound speed was decomposed into dark matter interacting with vacuum energy. In the synchronous gauge, we introduced an energy flow that was parallel to the four-velocity of the dark matter and obtained the evolution equations of growth rate for the dark matter and baryon. Then, we showed the effects on the CMB temperature and matter power spectra for the varied model parameter $\alpha$. When $\alpha$ was mean value or zero, from the power spectra, it was difficult to distinguish the IDCASS model from the $\Lambda$CDM model. However, due to using the $f\sigma_8(z)$ data set of large scale structure information, the evolutionary curves of $f\sigma_8(z)$ could break the degeneracy of the models.

Then, based on the MCMC method, a global fitting was performed on the decomposed model by adopting the CMB information from \textit{Planck}, BAO, SNIa, and RSD data sets. We obtained a tight constraint for the cosmological parameters. The results for three different cases were shown in Table \ref{tab:total-mean-table}. Obviously, the constraint with RSD data set is tighter than that without RSD data set, which means that the $f\sigma_8(z)$ data is very important to the cosmological constraint. With the data set of $f\sigma_8(z)$, the cosmic observational data sets all favor a small interaction which is up to the order of $10^{-4}$. It means that the IDCASS model and $\Lambda$CDM model undergo the similar background evolution behavior. Fortunately, the large scale structure information is a powerful tool to discriminate the dark energy models because the dynamical evolution would be different even if they had the same background evolution.

In future work, we will continue to study some other dark fluid, such as the generalized Chaplygin gas and modified Chaplygin gas model, by using the $f\sigma_8(z)$ data set. Moreover, if the entropy perturbation is considered, a negative adiabatic sound speed is favored, which is different from that of the pure adiabatic case. For the cosmic observations, we hope that some other data points of the growth rate can be found and released which could bring larger improvement into cosmological constraints.

\acknowledgements{L. Xu's work is supported in part by NSFC under the Grants No. 11275035 and "the Fundamental Research Funds for the Central Universities" under the Grants No. DUT13LK01.}

\end{document}